\documentclass[journal=jacsat,manuscript=article]{achemso}

\usepackage[version=3]{mhchem}
\usepackage{mathtools}
\usepackage{amsfonts,amsthm,bm}
\usepackage{xcolor}
\usepackage{soul}
\usepackage{siunitx}
\usepackage{comment}
\definecolor{myblue}{RGB}{0, 123, 167}

\usepackage{chapterbib}
\usepackage{hyperref}
\hypersetup{
    colorlinks=true,
    linkcolor=black,
    filecolor=blue,      
    urlcolor=blue,
    citecolor=black,
}
\makeatletter
\patchcmd{\acs@contact@details}{E}{*\,E}{}{}
\makeatother

\author{Ivan Riveros}
\author{Bin Zhang}
\affiliation{Department of Chemistry, Massachusetts Institute of Technology, Cambridge, MA 02139, USA}
\email{binz@mit.edu}

\title[SVGD Article]
{Collective-Variable Stein Variational Gradient Descent for Free Energy Surface Exploration}

\begin{document}

\section{Abstract}
Enhanced-sampling methods enable molecular simulations to access configurational regions that are difficult to explore with conventional dynamics, but their efficiency often deteriorates as the number of collective variables (CVs) increases. Here, we introduce an enhanced-sampling framework that couples Stein variational gradient descent (SVGD) with parallel restrained molecular dynamics simulations to facilitate exploration of higher-dimensional CV spaces. SVGD collectively transports an ensemble of interacting particles toward a target probability distribution by combining mean forces that guide particles toward low-free-energy regions with kernel-mediated repulsive interactions that maintain particle diversity. We validate the approach using an analytically solvable double-well model, followed by aqueous alanine dipeptide and tripeptide in two- and four-dimensional CV spaces, respectively. In each case, the particle ensemble reproduces the corresponding target distribution, and the reconstructed free-energy surfaces agree closely with analytical or umbrella-sampling references. For alanine dipeptide and tripeptide, CV-SVGD requires substantially less aggregate MD sampling than the corresponding umbrella-sampling reference calculations. These results demonstrate the potential of CV-SVGD as a flexible, parallelizable framework for enhanced sampling in multidimensional CV spaces. We additionally provide an open-source Python implementation to facilitate its adoption.

\clearpage
\newpage

\section{Introduction}
Free-energy calculations are central to the use of molecular simulations for studying conformational changes, molecular recognition, chemical reactions, and other processes governed by rare events \cite{pohorille_free_2007, frenkel_understanding_2023}. In principle, equilibrium properties can be obtained by sampling the Boltzmann distribution using molecular dynamics (MD) or Monte Carlo simulations. In practice, however, transitions between metastable states may occur on timescales far beyond those accessible to straightforward simulations, motivating the development of a broad range of enhanced-sampling methods \cite{torrie_nonphysical_1977,swendsen_replica_1986,piela_multiple-minima_1989,frantz_reducing_1990,e_marinari_simulated_1992,grubmuller_molecular_1994,huber_local_1994,grubmuller_predicting_1995,voter_hyperdynamics_1997,lee_new_1997,nakajima_multicanonical_1997,sugita_replica-exchange_1999,sorensen_temperature-accelerated_2000,hummer_free_2001,darve_calculating_2001,wang_efficient_2001,rosso_use_2002,laio_escaping_2002,bolhuis_transition_2002,park_free_2003,hamelberg_accelerated_2004,maragliano_temperature_2006,kim_statistical-temperature_2006,christ_enveloping_2007,abrams_efficient_2008,darve_adaptive_2008,gao_integrate-over-temperature_2008,pande_everything_2010,hansen_using_2010,ferguson_integrating_2011,chen_heating_2012,chodera_markov_2014,valsson_variational_2014,miao_gaussian_2015,yang_thermodynamics_2015,valsson_enhancing_2016,ribeiro_reweighted_2018,guo_adaptive_2018,perthold_accelerated_2018,rosa-raices_path-accelerated_2019,noe_boltzmann_2019,yang_enhanced_2019,ding_computing_2020,hartmann_infinite_2020,ding_deepbar_2021,sasmal_improved_2025}.

A common strategy underlying many enhanced-sampling approaches is to describe the molecular system using a reduced set of collective variables (CVs) that capture the slow or functionally relevant degrees of freedom. When an appropriate low-dimensional representation is available, methods such as umbrella sampling can substantially improve sampling efficiency \cite{torrie_nonphysical_1977,souaille_extension_2001,kumar_weighted_1992}. In practice, however, the difficulty of free-energy reconstruction generally increases with the number of CVs. This dimensionality dependence places strong pressure on identifying a compact representation, often consisting of only one or two coordinates, that retains the essential slow modes of the molecular process of interest.

Identifying such coordinates is itself a challenging problem. From a theoretical perspective, reaction coordinates related to the committor function provide an ideal description of transition progress,\cite{du_transition_1998} while recent machine-learning approaches offer increasingly sophisticated strategies for discovering low-dimensional representations from simulation data.\cite{rohrdanz_determination_2011,ferguson_integrating_2011,sidky_learning_2018,sidky_machine_2020,kang_committors_2026, ribeiro_reweighted_2018,peters_obtaining_2006,best_reaction_2005} In both cases, however, determining useful coordinates often requires representative sampling of the relevant metastable states and transition regions. This creates an inherent interdependence between CV discovery and enhanced sampling: effective CVs enable efficient exploration, yet substantial exploration may be required before those CVs can be reliably identified. As a result, the requirement to determine an optimal low-dimensional representation can itself become a bottleneck in free-energy calculations.

An alternative strategy is to reduce the dependence on aggressive dimensionality reduction by developing sampling methods that remain effective in richer CV spaces. Rather than requiring the essential molecular coordinates to be compressed into one or two variables before enhanced sampling begins, such approaches could permit sampling in a larger set of physically meaningful coordinates, with dimensionality reduction or mechanistic interpretation performed subsequently. Existing methods, including metadynamics \cite{laio_escaping_2002}, adaptive biasing approaches\cite{darve_adaptive_2008}, and temperature-accelerated molecular dynamics (TAMD)\cite{maragliano_temperature_2006}, can be extended to multidimensional CV spaces, but their efficiency can deteriorate as dimensionality increases. In particular, these trajectory-based enhanced-sampling methods still rely on the progressive exploration of CV space through biased or accelerated dynamics and may remain hindered by slow degrees of freedom or barriers not adequately represented by the chosen coordinates.

Here, we develop an enhanced-sampling framework based on Stein variational gradient descent (SVGD),\cite{liu_stein_2016} a particle-based variational inference method that transports an ensemble of particles toward a target probability distribution. We adapt SVGD to molecular simulations by defining particles in CV space and coupling each particle to a restrained MD simulation. The restrained simulations provide local thermodynamic information, while SVGD adaptively redistributes the particle locations so that the ensemble approaches the target equilibrium distribution.

This formulation provides a different mechanism for allocating sampling across CV space. Rather than relying on predefined sampling windows or on the progressive exploration of biased or accelerated trajectories, the locations of the restrained simulations are adaptively redistributed through a collective variational transport step. Consequently, exploration does not rely solely on spontaneous barrier-crossing events occurring on their natural timescales. The use of multiple particles also makes the framework compatible with parallel computation, because the restrained simulations associated with different particles can be propagated concurrently between SVGD updates.

We evaluate the method using a series of systems of increasing complexity and CV dimensionality to demonstrate the accuracy and efficiency of CV-SVGD in recovering equilibrium distributions and associated free-energy landscapes. To facilitate adoption of the approach, we additionally provide a Python implementation that couples SVGD updates with restrained molecular simulations and can be adapted to different molecular systems and CV definitions.

\section{Theory}

We explore an enhanced sampling strategy based on SVGD, in which an ensemble of particles is deterministically transported toward a target Boltzmann distribution. In the molecular simulation context, the particles can be interpreted as molecular configurations or, as described below, points in a lower-dimensional CV space. The SVGD update combines two effects: a force-like term that drives particles toward regions of high target probability, and a kernel-mediated repulsive term that maintains diversity among particles and prevents collapse into a single mode.

\subsection{Stein variational gradient descent}

SVGD formulates distributional approximation as a deterministic optimization problem. Given a target distribution $p(\mathbf{x})$, the goal is to evolve an initial trial distribution $q(\mathbf{x})$ so that it more closely matches $p(\mathbf{x})$ by minimizing the Kullback--Leibler (KL) divergence.

SVGD considers a smooth perturbation of the current particle positions,
\begin{equation}
    T(\mathbf{x}) = \mathbf{x} + \varepsilon \boldsymbol{\phi}(\mathbf{x}),
\end{equation}
where $\varepsilon$ is a small step size and $\boldsymbol{\phi}$ is a vector-valued velocity field. If $\mathbf{x}\sim q$, then the transformed variable $\mathbf{z}=T(\mathbf{x})$ follows a transformed distribution denoted by $q_T$. The agreement between $q_T$ and the target distribution $p$ is measured by
\begin{align}
    D_\mathrm{KL}(q_T\|p)
    &=
    \int d\mathbf{z}\,
    q_T(\mathbf{z})
    \log\frac{q_T(\mathbf{z})}{p(\mathbf{z})} \\
    &=
    \int d\mathbf{x}\,
    q(\mathbf{x})
    \log
    \frac{
    q(\mathbf{x})
    \left|\det \nabla_{\mathbf{x}}T(\mathbf{x})\right|^{-1}
    }{
    p(T(\mathbf{x}))
    } \\
    &=
    \mathbb{E}_{\mathbf{x}\sim q}
    \left[
    \log q(\mathbf{x})
    -
    \log\left|\det \nabla_{\mathbf{x}}T(\mathbf{x})\right|
    -
    \log p(T(\mathbf{x}))
    \right].
\end{align}

Taking the derivative of this KL divergence with respect to $\varepsilon$ at $\varepsilon=0$ gives
\begin{equation}
    \left.
    \frac{d}{d\varepsilon}
    D_\mathrm{KL}(q_T\|p)
    \right|_{\varepsilon=0}
    =
    -
    \mathbb{E}_{\mathbf{x}\sim q}
    \left[
    \operatorname{tr}
    \left(
    \mathcal{A}_p\boldsymbol{\phi}(\mathbf{x})
    \right)
    \right],
\end{equation}
where $\mathcal{A}_p$ is the Stein operator,
\begin{equation}
    \mathcal{A}_p\boldsymbol{\phi}(\mathbf{x})
    =
    \nabla_{\mathbf{x}}\log p(\mathbf{x})\,
    \boldsymbol{\phi}(\mathbf{x})^\top
    +
    \nabla_{\mathbf{x}}\boldsymbol{\phi}(\mathbf{x}).
\end{equation}
Thus, decreasing the KL divergence is equivalent to choosing a velocity field $\boldsymbol{\phi}$ that maximizes the expected Stein operator trace.

To obtain a tractable closed-form update, SVGD restricts $\boldsymbol{\phi}$ to a norm-bounded vector-valued reproducing kernel Hilbert space (RKHS), $\mathcal{H}^d$. Let $k(\mathbf{x},\mathbf{x}')$ be a positive-definite kernel with scalar RKHS $\mathcal{H}$. Each component $\phi_m$ of the vector field $\boldsymbol{\phi}=(\phi_1,\ldots,\phi_d)$ lies in $\mathcal{H}$ and satisfies the reproducing property,
\begin{equation}
    \phi_m(\mathbf{x})
    =
    \left\langle
    \phi_m(\cdot),
    k(\mathbf{x},\cdot)
    \right\rangle_{\mathcal{H}}.
\end{equation}
Under this RKHS constraint, the steepest descent direction of the KL divergence is
\begin{equation}
    \boldsymbol{\phi}_{q,p}^{\ast}(\cdot)
    =
    \mathbb{E}_{\mathbf{x}\sim q}
    \left[
    k(\mathbf{x},\cdot)\nabla_{\mathbf{x}}\log p(\mathbf{x})
    +
    \nabla_{\mathbf{x}}k(\mathbf{x},\cdot)
    \right].
\end{equation}

In practice, the expectation over $q$ is approximated using an ensemble of $n$ particles,
$\{\mathbf{x}_i^\ell\}_{i=1}^n$, at iteration $\ell$. The empirical SVGD update is
\begin{equation}
    \mathbf{x}_i^{\ell+1}
    =
    \mathbf{x}_i^\ell
    +
    \varepsilon_\ell
    \boldsymbol{\phi}^{\ast}(\mathbf{x}_i^\ell),
    \label{eq:svgd_config_update}
\end{equation}
with
\begin{equation}
    \boldsymbol{\phi}^{\ast}(\mathbf{x}_i^\ell)
    =
    \frac{1}{n}
    \sum_{j=1}^n
    \left[
    k(\mathbf{x}_j^\ell,\mathbf{x}_i^\ell)
    \nabla_{\mathbf{x}_j^\ell}\log p(\mathbf{x}_j^\ell)
    +
    \nabla_{\mathbf{x}_j^\ell}
    k(\mathbf{x}_j^\ell,\mathbf{x}_i^\ell)
    \right].
    \label{eq:svgd_empirical_update}
\end{equation}
For a Boltzmann target distribution,
\begin{equation}
    \nabla_{\mathbf{x}}\log p(\mathbf{x})
    =
    -\beta \nabla_{\mathbf{x}}U(\mathbf{x}).
\end{equation}
Therefore, the first term in Eq.~\ref{eq:svgd_empirical_update} moves particles along a kernel-smoothed thermodynamic force toward regions of lower potential energy and higher Boltzmann probability. The second term,
$\nabla_{\mathbf{x}_j}k(\mathbf{x}_j,\mathbf{x}_i)$, acts as a deterministic repulsive interaction among nearby particles, encouraging the ensemble to represent the full target distribution rather than collapsing into a single minimum.

\subsection{SVGD in collective-variable space}

Direct application of SVGD in the full molecular configuration space is challenging for high-dimensional molecular systems. First, the dimensionality of $\mathbf{x}$ can be very large, requiring many particles to represent the target distribution accurately. Second,  kernel methods become increasingly difficult in high dimensions because pairwise distances become less informative and kernel bandwidth selection becomes less robust \cite{scholkopf_learning_2001}.

To address these challenges, we apply SVGD in a lower-dimensional CV space. Let
\begin{equation}
    \boldsymbol{\xi}(\mathbf{x})
    =
    \left(
    \xi_1(\mathbf{x}),\ldots,\xi_N(\mathbf{x})
    \right)
\end{equation}
denote a set of $N$ CVs, and let $\boldsymbol{\zeta}$ denote a point in this CV space. The equilibrium probability density in CV space is
\begin{equation}
    p(\boldsymbol{\zeta})
    =
    \frac{1}{Z_{\zeta}}
    \exp[-\beta F(\boldsymbol{\zeta})],
\end{equation}
where $F(\boldsymbol{\zeta})$ is the free energy surface and $Z_{\zeta}$ is the corresponding normalization constant. Formally,
\begin{equation}
    F(\boldsymbol{\zeta})
    =
    -\beta^{-1}
    \log
    \left[
    \frac{1}{Z}
    \int d\mathbf{x}\,
    \exp[-\beta U(\mathbf{x})]\,
    \delta\left(\boldsymbol{\xi}(\mathbf{x})-\boldsymbol{\zeta}\right)
    \right],
    \label{eq:fes_definition}
\end{equation}
where $Z=\int d\mathbf{x}\,
    \exp[-\beta U(\mathbf{x})]$.

In CV space, the SVGD update has the same form as in configuration space (Eq.~\ref{eq:svgd_config_update}), but with $\boldsymbol{\zeta}$ replacing $\mathbf{x}$ and the CV Boltzmann distribution replacing the configurational Boltzmann distribution. The score function becomes
\begin{align}
    \nabla_{\boldsymbol{\zeta}}\log p(\boldsymbol{\zeta})
    =
    \nabla_{\boldsymbol{\zeta}}
    \log
    \left[
    \frac{1}{Z_{\zeta}}
    \exp[-\beta F(\boldsymbol{\zeta})]
    \right] 
   =
    -\beta
    \nabla_{\boldsymbol{\zeta}}F(\boldsymbol{\zeta}).
\end{align}
Thus, SVGD in CV space requires estimates of the mean force
$\nabla_{\boldsymbol{\zeta}}F(\boldsymbol{\zeta})$.


We estimate the mean force at a CV-space particle from restrained molecular simulations. For a harmonic restraint centered at $\boldsymbol{\zeta}$,
\begin{equation}
    U_{\kappa}(\mathbf{x};\boldsymbol{\zeta})
    =
    U(\mathbf{x})
    +
    \frac{\kappa}{2}
    \left\|
    \boldsymbol{\xi}(\mathbf{x})-\boldsymbol{\zeta}
    \right\|^2,
\end{equation}
the corresponding restrained free energy is
\begin{equation}
    F_{\kappa}(\boldsymbol{\zeta})
    =
    -\beta^{-1}
    \log
    \int d\mathbf{x}\,
    \exp
    \left[
    -\beta U(\mathbf{x})
    -
    \frac{\beta\kappa}{2}
    \left\|
    \boldsymbol{\xi}(\mathbf{x})-\boldsymbol{\zeta}
    \right\|^2
    \right].
    \label{eq:restrained_free_energy}
\end{equation}
Differentiating Eq.~\ref{eq:restrained_free_energy} with respect to the restraint center gives
\begin{equation}
    \nabla_{\boldsymbol{\zeta}}F_{\kappa}(\boldsymbol{\zeta})
    =
    \kappa
    \left\langle
    \boldsymbol{\zeta}
    -
    \boldsymbol{\xi}(\mathbf{x})
    \right\rangle_{\kappa,\boldsymbol{\zeta}},
    \label{eq:mean_force_estimator}
\end{equation}
where $\langle\cdot\rangle_{\kappa,\boldsymbol{\zeta}}$ denotes an equilibrium average in the restrained ensemble centered at $\boldsymbol{\zeta}$. In the stiff-spring limit, $F_{\kappa}(\boldsymbol{\zeta})$ approximates the underlying free energy surface up to a smoothing induced by the restraint width. Therefore, the restrained mean force in Eq.~\ref{eq:mean_force_estimator} provides a practical estimate of
$\nabla_{\boldsymbol{\zeta}}F(\boldsymbol{\zeta})$ for use in the CV-space SVGD update. This estimate was shown by \citet{maragliano_temperature_2006} to have an error of order $\mathcal{O}\left(\frac{1}{\beta\kappa}\right)$.

This formulation converts free-energy exploration into an interacting-particle optimization problem in CV space. Each SVGD particle represents a CV location at which local molecular simulations estimate the mean force, while the kernel interaction couples the particles so that they collectively explore and approximate the equilibrium distribution
$p(\boldsymbol{\zeta})\propto \exp[-\beta F(\boldsymbol{\zeta})]$.

\section{Results}
\subsection{SVGD as an Enhanced Sampling Method}

\begin{figure}[t]
    \centering
    \includegraphics[width=1.0\linewidth]{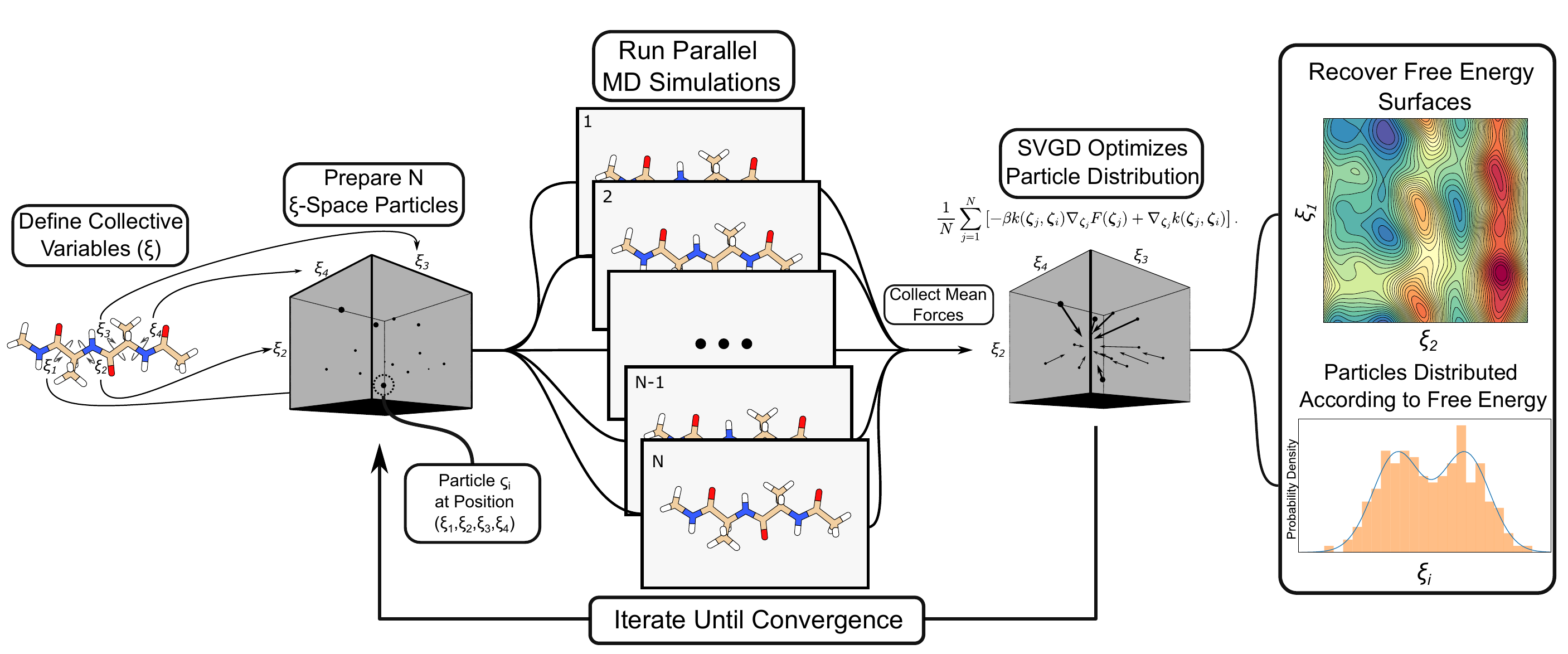}
\caption{
    \textbf{Workflow of CV-SVGD for adaptive exploration and free-energy reconstruction.}
    A set of particles is initialized in a chosen CV space, with each particle associated with a restrained molecular simulation centered at its current CV position. The restrained simulations are propagated in parallel to estimate local mean forces, which are then combined through the SVGD update to redistribute the particle ensemble toward a chosen target distribution in CV space. This cycle is repeated until convergence. The mean-force information collected at the particle locations is subsequently used to reconstruct the multidimensional free-energy surface.
}
    \label{fig:workflow}
\end{figure}

The theory above provides a direct route for applying SVGD to molecular free-energy calculations. In CV-SVGD, each particle represents a location in CV space and is associated with a restrained MD simulation centered at that location. At each iteration, these simulations provide local estimates of the mean force, which are used in the SVGD update to collectively redistribute the particles in CV space. The restraints then guide the underlying molecular configurations toward the updated particle positions, and the procedure is repeated iteratively (Figure~\ref{fig:workflow}).

Because practical SVGD calculations can use a relatively small number of particles, the particle positions themselves may provide only a sparse representation of the equilibrium distribution and may therefore be insufficient for directly estimating a detailed free-energy surface. Instead, we reconstruct the free-energy surface from the mean forces evaluated at the particle locations. In this way, SVGD determines where molecular sampling is performed, while the resulting mean-force information is used for free-energy reconstruction.

We implemented CV-SVGD in the Python package CVSVGD. The implementation automates the coupling between restrained molecular simulations, mean-force estimation, and SVGD particle updates, and supports parallel execution of the restrained simulations associated with different particles. The workflow can be applied to arbitrary choices of CVs and iterated for a prescribed number of steps or until the particle distribution has converged.

\subsection{Validation of CV-SVGD on an Analytically Solvable Model}

We first validate CV-SVGD using an analytically solvable three-dimensional double-well model. The potential is defined as
\begin{equation}
    U(\mathbf{x})=-\ln\left[
    \frac{1}{2}\left(\frac{k_1}{2\pi}\right)^{3/2}
    e^{-\frac{k_1}{2}\left\|\mathbf{x}-\mathbf{x}_1\right\|^2}
    +
    \frac{1}{2}\left(\frac{k_2}{2\pi}\right)^{3/2}
    e^{-\frac{k_2}{2}\left\|\mathbf{x}-\mathbf{x}_2\right\|^2}
    \right],
    \label{eq:toy_model_potential}
\end{equation}
where $\mathbf{x}=(x,y,z)$, $k_1$ and $k_2$ determine the widths of the two wells, and $\mathbf{x}_1$ and $\mathbf{x}_2$ denote their centers. We consider two parameterizations of this model, with different values of $(k_1,k_2)$, that produce relatively low and high free-energy barriers between the two basins (see Methods).

To mimic the application of CV-SVGD to a molecular system, we treat $(x,y)$ as the CVs and integrate over $z$, thereby defining a two-dimensional CV space. We use 1000 SVGD particles and evaluate the mean forces analytically using Eq.~\ref{eq:mean_force_toy}.

For the low-barrier system, the SVGD particles are evolved at the physical inverse temperature, $\beta=1$, and therefore directly target the equilibrium probability distribution. For the high-barrier system, we introduce a simple temperature-scaling strategy to improve coverage of CV space. The mean force $\nabla F(x,y)$ continues to correspond to the free-energy surface of the original system at $\beta=1$, but its contribution to the SVGD update is scaled by a smaller effective inverse temperature, $\beta_H=0.024$. The particles therefore evolve toward the flattened distribution
\begin{equation}
    p_H(x,y) \propto \exp[-\beta_H F(x,y)],
    \label{eq:high_temp_toy_p}
\end{equation}
which reduces the contrast between low- and high-free-energy regions and prevents the particle ensemble from becoming overly concentrated within the free-energy minima. This strategy is conceptually analogous to the temperature acceleration used in AFED and TAMD.\cite{maragliano_temperature_2006,rosso_use_2002,abrams_efficient_2008} Importantly, changing $\beta_H$ alters only the target distribution used to guide placement of the SVGD particles; it does not modify the underlying free-energy surface or the mean forces used for its reconstruction.

\begin{figure}[t]
    \centering
    \includegraphics[width=1.0\linewidth]{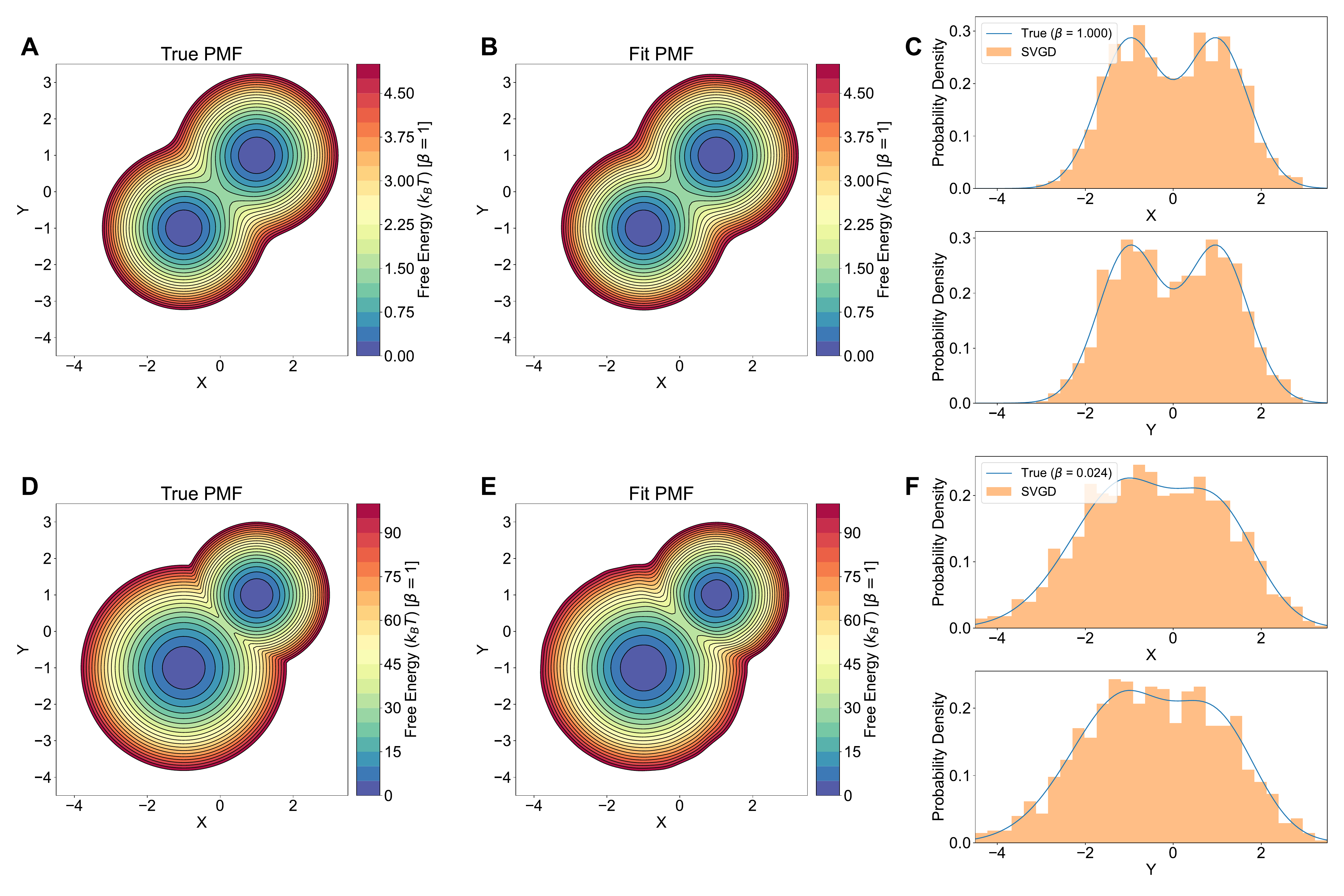}
\caption{
    \textbf{Validation of CV-SVGD on an analytically solvable double-well model.}
    (A,D) Analytical free-energy surfaces in the $(x,y)$ CV space for the low-barrier (A) and high-barrier (D) systems.
    (B,E) Free-energy surfaces reconstructed from mean forces evaluated at the final CV-SVGD particle locations for the low-barrier (B) and high-barrier (E) systems.
    (C,F) One-dimensional marginal distributions of the CV-SVGD particles compared with the corresponding analytical target distributions.
}
    \label{fig:toy_model}
\end{figure}

After convergence, the final SVGD particle distributions closely follow the corresponding analytical target distributions (Figure~\ref{fig:toy_model}C,F). For the low-barrier system, the particles reproduce the equilibrium distribution at $\beta=1$. For the high-barrier system, they instead sample the deliberately flattened distribution defined by $\beta_H=0.024$, again showing close agreement with the analytical result.

We next use the particle locations as centers at which mean-force information is available and reconstruct the free-energy surface by fitting the mean forces with Gaussian radial-basis functions.\cite{maragliano_single-sweep_2008} For both the low- and high-barrier systems, the reconstructed surfaces closely reproduce the corresponding analytical free-energy surfaces (Figure~\ref{fig:toy_model}A,B,D,E).

Together, these results demonstrate that CV-SVGD can use local mean-force information both to adaptively distribute sampling across CV space and to recover the underlying free-energy surface, even when the particle distribution is generated using an accelerated effective temperature.

\subsection{Aqueous-Phase Alanine Dipeptide}

\begin{figure}[t]
    \centering
    \includegraphics[width=1.0\linewidth]{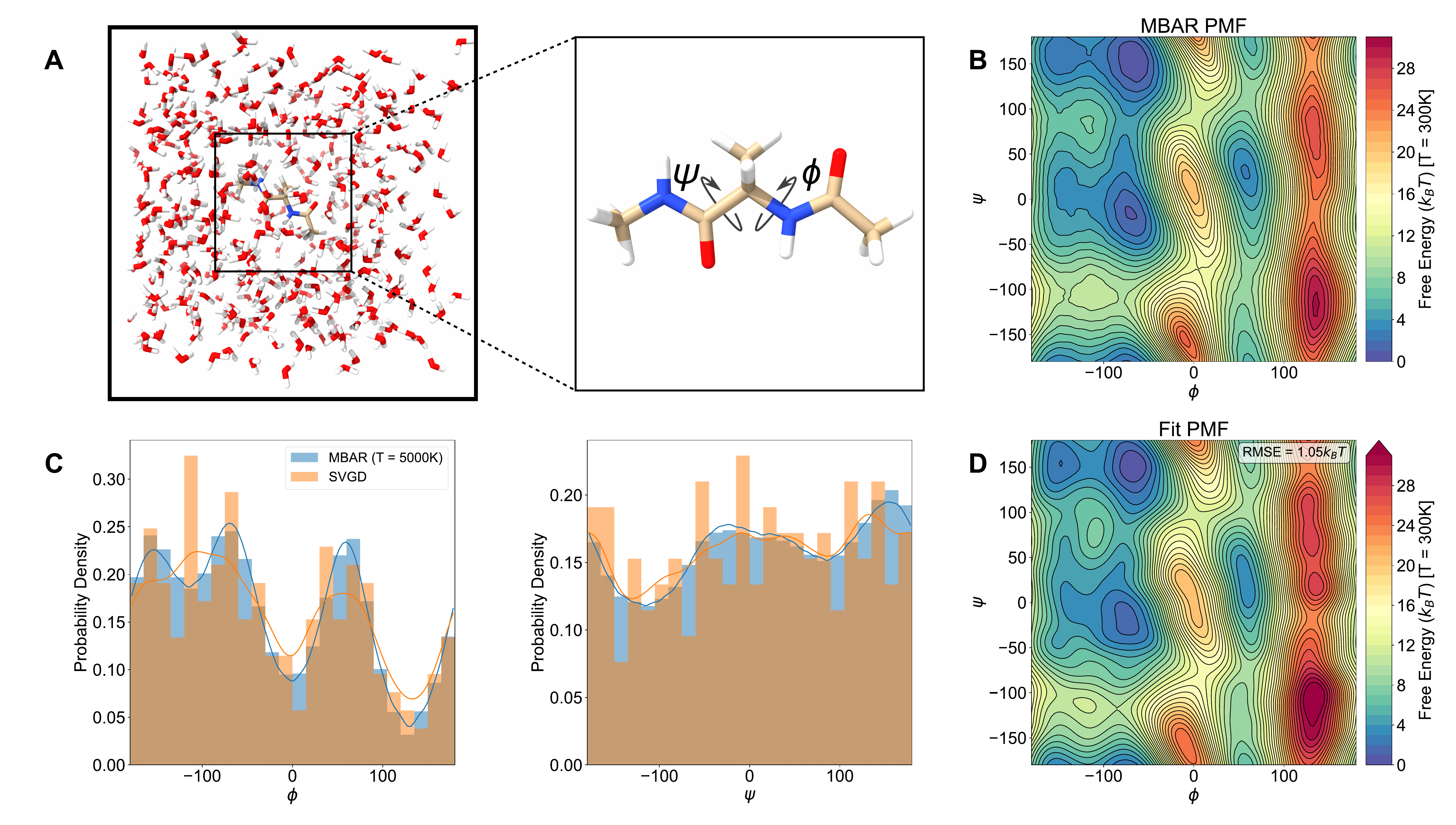}
\caption{\textbf{CV-SVGD recovers the free-energy landscape of aqueous-phase alanine dipeptide.}
(A) Alanine dipeptide and the backbone dihedral angles $\phi$ and $\psi$ used to define the CV space.
(B) Reference free-energy surface obtained from two-dimensional umbrella sampling and MBAR reweighting.
(C) Comparison of one-dimensional marginal distributions of the final SVGD particles with the corresponding temperature-scaled distribution derived from the umbrella-sampling reference.
(D) Free-energy surface reconstructed from CV-SVGD mean forces. The root-mean-square error (RMSE) relative to the umbrella-sampling surface is reported in the upper right. The triangular extension of the color bar denotes free-energy values above the displayed range of the umbrella-sampling reference.}
    \label{fig:alanine_dipeptide}
\end{figure}

We next apply CV-SVGD to aqueous-phase alanine dipeptide as a proof of concept, using the backbone Ramachandran dihedral angles $\phi$ and $\psi$ to define the two-dimensional CV space (Figure~\ref{fig:alanine_dipeptide}A). Alanine dipeptide is a prototypical benchmark for enhanced-sampling methods,\cite{rosso_mapping_2005,bartels_multidimensional_1997,ensing_metadynamics_2006,maragliano_single-sweep_2008} because its conformational free-energy landscape is well characterized and can be obtained accurately using established approaches. We therefore constructed a reference free-energy surface in $(\phi,\psi)$ space using umbrella-sampling simulations.\cite{torrie_nonphysical_1977} Unbiased statistical weights were obtained by solving the multistate Bennett acceptance ratio (MBAR) equations\cite{shirts_statistically_2008} using FastMBAR,\cite{ding_fast_2019} and the resulting reference PMF is shown in Figure~\ref{fig:alanine_dipeptide}B.

As in the high-barrier toy model, we used temperature scaling to improve the spatial coverage of the SVGD particles. The restrained MD simulations used to estimate mean forces were performed at the physical temperature of 300~K, while the SVGD particle dynamics used an effective temperature of 5000~K. We propagated 200 particles for 40 SVGD iterations, using 1~ps of restrained MD per particle at each iteration to estimate the local mean force. After the final iteration, the particle locations were fixed and additional 50~ps restrained simulations were performed at each location to obtain more accurate mean-force estimates for free-energy reconstruction.

The final SVGD particle distribution is consistent with the temperature-scaled target distribution derived from the umbrella-sampling reference (Figure~\ref{fig:alanine_dipeptide}C). The one-dimensional marginal distributions reproduce the major features of the reference distribution, with residual deviations that likely arise from finite particle number and histogram discretization. More importantly, the free-energy surface reconstructed from the mean-force data closely reproduces the umbrella-sampling result (Figure~\ref{fig:alanine_dipeptide}D). CV-SVGD recovers the locations and relative organization of the major metastable basins and the intervening barriers, with an overall root-mean-square error of 1.05~$k_\mathrm{B}T$ relative to the umbrella-sampling PMF. Although the production calculation was propagated for 40 SVGD iterations, convergence analysis indicates that comparable accuracy may already be achieved after approximately 20 iterations (Figure~\ref{fig:si_alanine_dipeptide_conv}).

The complete CV-SVGD calculation required approximately 18~ns of aggregate restrained-MD sampling. By comparison, the reference umbrella-sampling calculation consisted of 625 windows, each simulated for 500~ps, corresponding to 312.5~ns of aggregate MD sampling. 
Thus, for this benchmark, CV-SVGD reproduced the reference free-energy landscape using substantially less total MD sampling. 

\subsection{Aqueous-Phase Alanine Tripeptide}

\begin{figure}[H]
    \centering
    \includegraphics[width=1.0\linewidth]{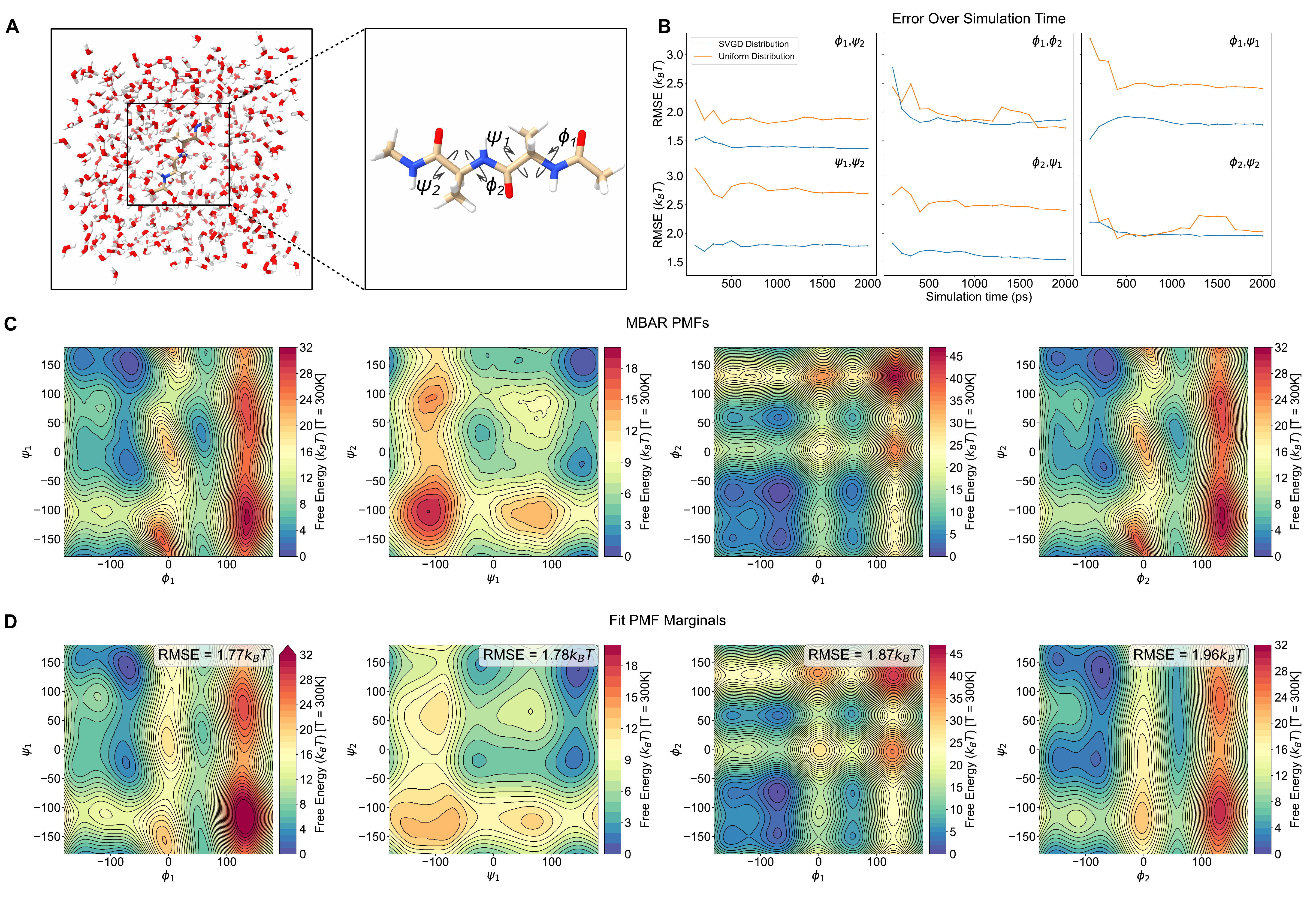}
    \caption{\textbf{CV-SVGD recovers the four-dimensional free-energy landscape of aqueous-phase alanine tripeptide.}
    (A) Alanine tripeptide and the four backbone dihedral angles used to define the CV space.
    (B) Accuracy of free-energy profiles reconstructed using mean forces evaluated at particle locations drawn from the converged SVGD distribution, compared with profiles obtained using uniformly distributed particle locations. Agreement with the US reference is quantified by the RMSE.
    (C) MBAR-reweighted free-energy surfaces obtained from six independent two-dimensional umbrella-sampling calculations, corresponding to all unique pairwise projections of the four-dimensional CV space.
    (D) Corresponding two-dimensional projections of the four-dimensional free-energy surface reconstructed from CV-SVGD mean forces. RMSE values relative to the umbrella-sampling references are reported in the upper right of each panel. Triangular extensions of the color bars denote free-energy values above the displayed range of the corresponding umbrella-sampling surface.}
    \label{fig:alanine_tripeptide}
\end{figure}

To further test CV-SVGD in a higher-dimensional setting, we applied the method to aqueous-phase alanine tripeptide using the four backbone Ramachandran dihedral angles $(\phi_1,\psi_1,\phi_2,\psi_2)$ as CVs (Figure~\ref{fig:alanine_tripeptide}A). Direct construction of a high-resolution four-dimensional reference free-energy surface using conventional umbrella sampling would be computationally demanding. For example, a grid containing 25 windows along each coordinate would require $25^4=390{,}625$ umbrella windows. We therefore constructed reference data using six independent two-dimensional umbrella-sampling calculations, corresponding to all unique pairwise projections of the four-dimensional CV space.

As for alanine dipeptide, we used temperature scaling to improve coverage of CV space. The restrained MD simulations used to estimate mean forces were performed at 300~K, whereas the SVGD particle dynamics used an effective temperature of 10,000~K. An ensemble of 200 particles was propagated for 200 SVGD iterations, with 1~ps of restrained MD per particle and per iteration used to estimate the local mean forces. After the final iteration, longer restrained simulations were performed at the converged particle locations to obtain more accurate mean-force estimates for reconstruction of the four-dimensional free-energy surface.

Because a direct four-dimensional umbrella-sampling reference was not available, we assessed the reconstructed CV-SVGD surface through its six unique two-dimensional projections. The projected CV-SVGD free-energy surfaces reproduce the major metastable basins and intervening barriers observed in the corresponding umbrella-sampling references (Figure~\ref{fig:alanine_tripeptide}C,D). Some projections exhibit smoother features and modest underestimation of barrier heights. These differences likely reflect a combination of finite particle density, statistical uncertainty in the estimated mean forces, and smoothing introduced during the radial-basis-function reconstruction and subsequent marginalization.

\begin{figure}[t]
    \centering
    \includegraphics[width=1.0\linewidth]{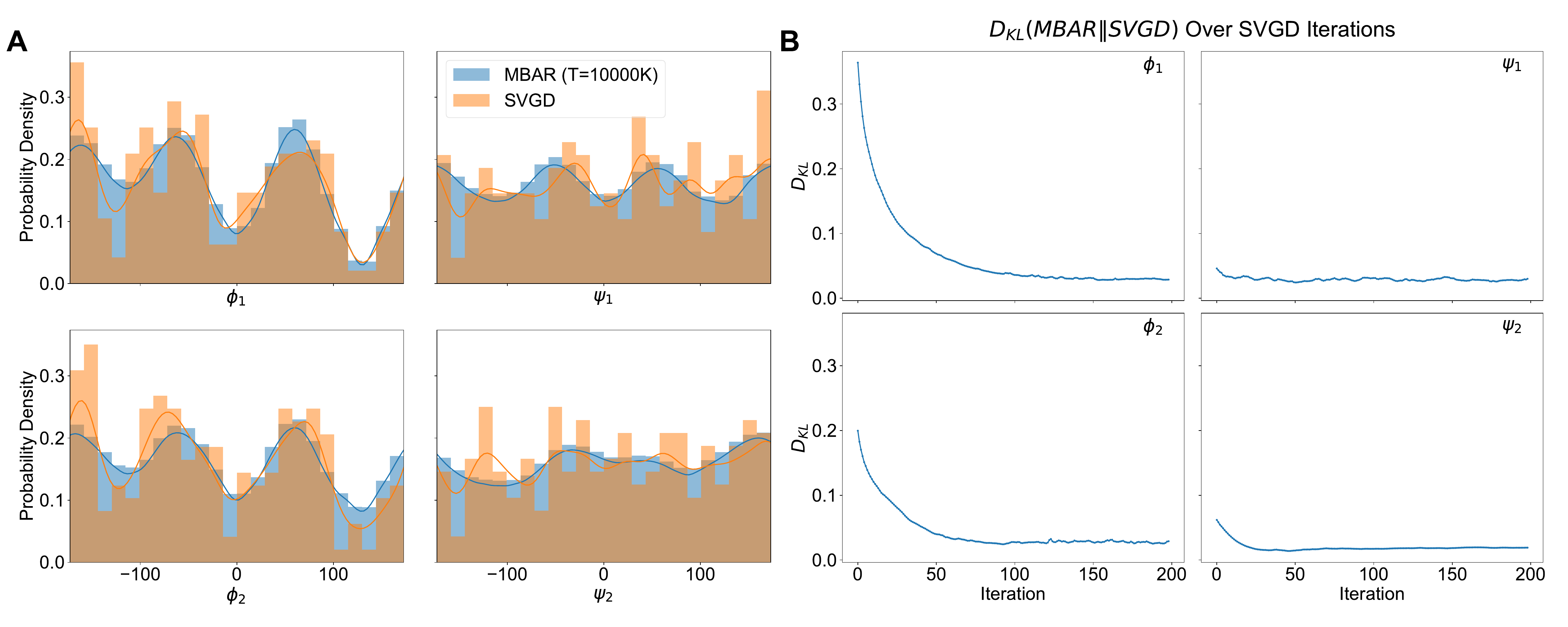}
    \caption{\textbf{CV-SVGD reproduces the reference marginal distributions of aqueous-phase alanine tripeptide.}
    (A) Comparison of the one-dimensional marginal distributions of the final SVGD particles with the corresponding temperature-scaled distributions obtained from MBAR-reweighted umbrella-sampling data. Kernel-density estimates (KDEs) are shown for both distributions.
    (B) Kullback--Leibler divergence, $D_{\mathrm{KL}}$, between the SVGD and umbrella-sampling KDEs as a function of SVGD iteration. }
    \label{fig:alanine_tripeptide_conv}
\end{figure}

The particle distributions provide an independent assessment of convergence that does not depend on reconstruction of the free-energy surface. As shown in Figure~\ref{fig:alanine_tripeptide_conv}A, the one-dimensional marginal distributions of the converged SVGD particles closely reproduce the corresponding temperature-adjusted distributions obtained from the MBAR-reweighted umbrella-sampling data. The Kullback--Leibler divergences between kernel-density estimates of the SVGD and reference distributions decrease rapidly and largely plateau within approximately 100 iterations for all four CVs (Figure~\ref{fig:alanine_tripeptide_conv}B). This agreement suggests that part of the residual discrepancy in the reconstructed free-energy surfaces arises from finite mean-force sampling and fitting rather than from failure of the SVGD particle ensemble to approximate the target distribution.

We next examined whether the adaptive placement of particles by SVGD improves free-energy reconstruction relative to a non-adaptive distribution of sampling locations. Figure~\ref{fig:alanine_tripeptide}B compares free-energy profiles reconstructed from mean forces evaluated at the converged SVGD particle positions with those obtained using particle positions distributed uniformly throughout the four-dimensional dihedral-angle space. The SVGD-distributed force centers consistently produce better overall agreement with the umbrella-sampling references. This comparison highlights an important role of the SVGD dynamics beyond distributional sampling: by preferentially allocating particles to thermodynamically relevant regions while maintaining coverage through kernel-mediated repulsion, CV-SVGD provides a more informative set of locations for subsequent mean-force reconstruction.

The alanine-tripeptide calculation also illustrates a distinction between convergence of the SVGD particle distribution and convergence of the reconstructed free-energy surface. Short, 1-ps restrained simulations were sufficient for the SVGD ensemble to approach the reference marginal distributions, whereas substantially longer simulations at the final particle locations were required to obtain mean forces accurate enough for quantitative free-energy reconstruction. Thus, the particle transport itself appears relatively tolerant of noisy local force estimates, whereas accurate reconstruction of a multidimensional free-energy surface places more stringent demands on the precision of the mean forces. Increasing the number of particles provides an additional route to improving the reconstruction: simulations with 500 particles reduce the RMSE of the projected free-energy surfaces to approximately $1.0\,k_\mathrm{B}T$ (Figure~\ref{fig:si_alanine_tripeptide}).

In terms of aggregate simulation cost, the 200-particle SVGD propagation requires 40ns of restrained-MD sampling. Including 200ns of additional sampling at each final particle location gives a total cost of 0.24$\mu$s. For comparison, the six two-dimensional umbrella-sampling reference calculations require approximately $0.375~\mu\mathrm{s}$ of aggregate simulation time to reach errors below $2.0\,k_\mathrm{B}T$ (Figure~\ref{fig:si_us_convergence}). Importantly, however, these umbrella calculations provide only six two-dimensional projections. A direct four-dimensional umbrella-sampling calculation with 25 windows per dimension would require 390,625 windows; even at 100~ps per window, this corresponds to approximately $39~\mu\mathrm{s}$ of aggregate simulation time. CV-SVGD therefore becomes increasingly attractive as the dimensionality of the target CV space grows, because the number of restrained simulations is determined by the number of particles rather than by a tensor-product grid over all CVs.

\section{Conclusions and Discussion}

In this work, we introduced CV-SVGD, an enhanced-sampling framework that combines SVGD with restrained MD to explore free-energy landscapes in CV space. By using local mean-force estimates to collectively transport an ensemble of CV-space particles, the method provides an alternative to conventional trajectory-based exploration while retaining an explicit connection to molecular configurations. Across a series of model and molecular systems, CV-SVGD accurately reproduced equilibrium distributions and free-energy surfaces, including multidimensional landscapes.

CV-SVGD shares several conceptual features with TAMD and related approaches \cite{abrams_efficient_2008,chen_heating_2012}. In both cases, the evolution of CVs is driven by mean forces estimated from restrained molecular simulations, and an elevated effective temperature can be used to facilitate exploration of CV space. The principal distinction is that TAMD propagates auxiliary CVs through a continuous dynamical trajectory, whereas CV-SVGD evolves an interacting ensemble of particles through a collective variational transport step. As a result, exploration in CV-SVGD is not governed solely by sequential barrier-crossing events in CV space. Information from the entire particle ensemble contributes to each update, while the underlying restrained molecular simulations can be carried out concurrently. This combination of collective transport and parallel sampling may become increasingly advantageous as the dimensionality and complexity of the CV space increase.

We anticipate that this capability may be particularly useful for systems in which the relevant slow degrees of freedom are not known a priori. Examples include biomolecular condensates \cite{latham_microphase_2025,latham_stability_2022,liu_toward_2025,wang_nonspecific_2024}, chromatin organization \cite{liu_chromatin_2022,qiu_nucleosome_2026,lin_cooperative_2021,paggi_euchromatin_2026,paggi_toward_2025}, and other heterogeneous molecular assemblies whose behavior may depend on many coupled structural and compositional coordinates \cite{srinivasan_ligand-induced_2022}. In such systems, identifying a minimal set of optimal CVs can itself be difficult. A sampling framework that remains effective in a richer CV representation could reduce the need to determine this low-dimensional description before enhanced sampling begins. In the present study, we explicitly reconstructed free-energy surfaces containing up to four CVs. This limit was dictated primarily by the difficulty of constructing and validating reliable high-dimensional reference free-energy surfaces rather than by a demonstrated breakdown of CV-SVGD. We plan to determine the practical dimensional limits of the method in future studies.

Several aspects of the present implementation could be improved further. First, we reconstruct the free-energy surface exclusively from mean-force estimates. Accurate mean forces can require substantial sampling at each particle location and may therefore become a significant component of the computational cost. The restrained trajectories, however, contain additional statistical information beyond their mean forces. Combining force-based reconstruction with configuration-based estimators, for example through MBAR-like reweighting or related multistate approaches, could make more complete use of the generated simulation data and potentially reduce the amount of sampling required to achieve a given free-energy accuracy.

Second, the efficiency of CV-SVGD depends on how the particles are distributed across CV space. In the present implementation, we used an elevated effective temperature to flatten the target distribution and prevent excessive concentration of particles within low-free-energy basins. More adaptive strategies may provide better control over particle placement. For example, the free-energy surface accumulated during successive SVGD iterations could be used to construct an evolving bias potential, analogous in spirit to adaptive-biasing approaches \cite{darve_adaptive_2008,guo_adaptive_2018,wang_efficient_2022}. The particle dynamics could then be updated using this progressively refined estimate, allowing sampling effort to be redistributed toward poorly explored regions while retaining the information accumulated in previous iterations. Such an approach could further improve both coverage of high-dimensional CV spaces and the statistical efficiency of free-energy reconstruction.


\section{Methods}
\subsection{General SVGD Protocol}

All CV-SVGD calculations were performed using our Python implementation, CVSVGD. For the molecular systems, restrained MD simulations were performed using OpenMM\cite{eastman_openmm_2024} to estimate the local mean forces required for the SVGD updates. Because the restrained simulations associated with different particles are independent between SVGD updates, they can be executed concurrently. We used the Parsl Python library\cite{babuji2019Parsl} to manage parallel execution across distributed computing resources.

The SVGD particle positions were updated using the Adam optimizer\cite{kingma_adam_2017} implemented in PyTorch.\cite{paszke_pytorch_2019} The interactions between particles were evaluated using the inverse multiquadric (IMQ) kernel, which has favorable convergence properties for certain classes of high-dimensional target distributions.\cite{gorham_measuring_2017} The kernel is defined as
\begin{equation}
    k(\mathbf{x}_i,\mathbf{x}_j)
    =
    \left(
    1+\frac{\|\mathbf{x}_i-\mathbf{x}_j\|^2}{h}
    \right)^{-1/2},
\end{equation}
where the bandwidth parameter $h$ was determined from the median pairwise particle distance at each iteration. The Adam learning rate was set to 0.6 for all calculations, with additional optimizer parameters specified below.


Following convergence of the SVGD particle ensemble, additional restrained simulations were performed at the final particle locations to obtain more accurate mean-force estimates for free-energy reconstruction. The resulting mean forces were fitted using radial-basis functions (RBFs), following the approach of Maragliano et al.\cite{maragliano_single-sweep_2008} Details of the RBF fitting procedure are provided in the Supporting Information.

\subsection{Two-dimensional Toy System}

We validated CV-SVGD using an analytically solvable three-dimensional double-well potential, defined in Eq.~\ref{eq:toy_model_potential}. Taking $(x,y)$ as the CVs and integrating out the $z$ coordinate, the corresponding free-energy gradient is
\begin{equation}
    \nabla_{x,y}F(x,y)
    =
    -\beta^{-1}\nabla_{x,y}
    \ln\int dz\,e^{-\beta U(\mathbf{x})}.
\end{equation}
At $\beta=1$, this expression can be evaluated analytically:
\begin{equation}
\begin{aligned}
    \nabla_{x,y}F(x,y)
    &=
    \frac{
    k_1^2(\mathbf{s}-\mathbf{s}_1)
    e^{-\frac{k_1}{2}\|\mathbf{s}-\mathbf{s}_1\|^2}
    +
    k_2^2(\mathbf{s}-\mathbf{s}_2)
    e^{-\frac{k_2}{2}\|\mathbf{s}-\mathbf{s}_2\|^2}
    }{
    k_1e^{-\frac{k_1}{2}\|\mathbf{s}-\mathbf{s}_1\|^2}
    +
    k_2e^{-\frac{k_2}{2}\|\mathbf{s}-\mathbf{s}_2\|^2}
    },
\end{aligned}
\label{eq:mean_force_toy}
\end{equation}
where $\mathbf{s}=(x,y)$ and
$\mathbf{s}_1=(x_1,y_1)$ and $\mathbf{s}_2=(x_2,y_2)$
denote the projected well centers.

We considered two parameterizations with different barrier heights. The low-barrier system used $k_1=k_2=2$, whereas the high-barrier system used $k_1=50$ and $k_2=25$. For both systems, the well centers were fixed at $\mathbf{x}_1=(1,1,1)$ and $\mathbf{x}_2=(-1,-1,-1)$.

At $\beta=1$, the analytical equilibrium distribution in the two-dimensional CV space is
\begin{equation}
    P(x,y)=
    \frac{1}{2}\left(\frac{k_1}{2\pi}\right)
    e^{-\frac{k_1}{2}\|\mathbf{s}-\mathbf{s}_1\|^2} +
    \frac{1}{2}\left(\frac{k_2}{2\pi}\right)
    e^{-\frac{k_2}{2}\|\mathbf{s}-\mathbf{s}_2\|^2}.
\end{equation}

For the low-barrier system, SVGD was performed at $\beta=1$, targeting the equilibrium distribution $P(x,y)$. For the high-barrier system, we employed temperature scaling with an effective inverse temperature $\beta_H=0.024$, while retaining the free-energy gradients evaluated at $\beta=1$. The corresponding SVGD target distribution is
\begin{equation}
    P_{\mathrm{SVGD}}(x,y)
    =
    \frac{P(x,y)^{\beta_H}}
    {\displaystyle\iint dx\,dy\,P(x,y)^{\beta_H}},
    \label{eq:high_temp_toy_prob}
\end{equation}
where $\beta_H$ controls the degree of flattening of the target distribution without modifying the underlying free-energy surface.

For both systems, 1000 SVGD particles were initialized uniformly over the region $x,y\in[-5,5]$ and propagated for 150 iterations using the analytical free-energy gradients. The particle positions obtained after the final iteration were used as centers for RBF fitting to reconstruct the free-energy surfaces.

\subsection{Alanine Dipeptide}

\subsubsection{CV-SVGD Simulations}

For aqueous-phase alanine dipeptide, we initialized 200 SVGD particles uniformly over the two-dimensional Ramachandran dihedral-angle space $(\phi,\psi)$. Initial molecular configurations were generated using tleap in AmberTools\cite{case_ambertools_2023}. Before initiating the SVGD updates, each configuration was subjected to a 10~ps restrained MD simulation to guide the dihedral angles toward their assigned particle positions.

All MD simulations were performed using OpenMM with the ff14SB protein force field\cite{maier_ff14sb_2015} and the TIP3P-FB explicit water model.\cite{jorgensen_comparison_1983,wang_building_2014} The system was solvated with 423 water molecules in a cubic periodic box of side length 2.4~nm. Long-range electrostatic interactions were treated using the particle mesh Ewald method,\cite{darden_particle_1993} with a nonbonded cutoff of 1.0~nm. Simulations were propagated using the Langevin middle integrator at 300~K, with a friction coefficient of 5~ps$^{-1}$ and a time step of 1~fs. Molecular configurations were recorded every 0.01~ps for mean-force estimation.

To constrain molecular configurations near the corresponding SVGD particle positions in CV space, we applied a harmonic restraining potential of the form
\begin{equation}
    R(\mathbf{x};\phi_i,\psi_i)
    =
    \frac{\kappa}{2}\left[
    (\phi(\mathbf{x})-\phi_i)^2
    +
    (\psi(\mathbf{x})-\psi_i)^2
    \right],
\end{equation}
where $\mathbf{x}$ denotes the molecular coordinates, $\phi(\mathbf{x})$ and $\psi(\mathbf{x})$ are the instantaneous backbone dihedral angles, and $(\phi_i,\psi_i)$ specifies the position of the $i$th SVGD particle in CV space. The restraint force constant was set to $\kappa=2500~\mathrm{kJ\,mol^{-1}\,rad^{-2}}$.

The SVGD particle ensemble was optimized at an effective temperature of 5000~K, while the restrained MD simulations used for mean-force estimation remained at 300~K. Particle positions were updated using the Adam optimizer with $\beta_1=0.6$, $\beta_2=0.999$, $\epsilon=0.1$, and a learning rate of 0.6.

Following each SVGD update, an additional 0.1~ps restrained MD simulation was performed to allow the molecular configurations to relax toward the updated particle positions. This intermediate relaxation step reduced the occurrence of excessively large forces arising from abrupt changes in restraint centers, allowing the use of larger SVGD updates without compromising numerical stability.

After the final SVGD iteration, the particle positions were fixed and an additional 50~ps restrained MD simulation was performed at each location to obtain more accurate mean-force estimates for free-energy reconstruction. The free-energy surface was reconstructed by RBF fitting using von Mises functions, which account for the periodicity of the dihedral angles. The one-dimensional von Mises basis function takes the form
\begin{equation}
    k_{\mathrm{VM}}(\theta)
    =
    \exp\left[\lambda\cos(\theta)\right],
\end{equation}
where $\lambda$ controls the width of the basis function. These periodic basis functions were used in place of Gaussian functions for fitting the free-energy surface in Ramachandran space.

\subsubsection{Umbrella Sampling Simulations}

Reference free-energy surfaces were obtained using two-dimensional umbrella sampling in the $(\phi,\psi)$ dihedral-angle space. The molecular model and simulation parameters were identical to those used for CV-SVGD, except that the restraint force constant was reduced to $\kappa=500~\mathrm{kJ\,mol^{-1}\,rad^{-2}}$ and the integration time step was increased to 2~fs.

Each dihedral angle was discretized into 25 equally spaced umbrella windows spanning $[-\pi,\pi)$, yielding a total of $25\times25=625$ windows. Each window was simulated for 0.5~ns, with configurations recorded every 1~ps.


\subsection{Alanine Tripeptide}

\subsubsection{CV-SVGD Simulations}

For aqueous-phase alanine tripeptide, we initialized 200 SVGD particles uniformly over the four-dimensional backbone dihedral-angle space $(\phi_1,\psi_1,\phi_2,\psi_2)$. The SVGD particle ensemble was propagated for 200 iterations using an effective temperature of 10,000~K, while the restrained MD simulations used for mean-force estimation were performed at 300~K. Each restrained simulation was 1~ps in duration.

The molecular system was solvated with 444 water molecules in a cubic periodic box of side length 2.5~nm. All other molecular simulation parameters, including the force field, water model, integrator, and restraint protocol, were identical to those used for alanine dipeptide.

Following the final SVGD iteration, additional 2~ns restrained MD simulations were performed at each particle location to reduce statistical noise in the mean-force estimates. These refined mean forces were subsequently used to reconstruct the four-dimensional free-energy surface through radial-basis-function fitting with von Mises kernels.

\subsubsection{Umbrella Sampling Simulations}

To validate the four-dimensional free-energy surface obtained from CV-SVGD, we performed six independent sets of two-dimensional umbrella-sampling simulations, corresponding to all unique pairs of the four backbone dihedral angles. 

The molecular model and simulation parameters were identical to those used for CV-SVGD, except that the restraint force constant was reduced to $\kappa=250~\mathrm{kJ\,mol^{-1}\,rad^{-2}}$ and the integration time step was increased to 2~fs. Each dihedral angle was discretized into 25 equally spaced umbrella windows spanning $[-\pi,\pi)$, yielding $25\times25=625$ windows per CV pair and 3,750 windows across all six umbrella-sampling calculations. Each window was simulated for 0.5~ns, with configurations recorded every 1~ps.

\begin{acknowledgement}
    This work was supported by the National Institutes of Health (Grant R35GM1335w80). I.R. was the recipient of the National Science Foundation Graduate Research Fellowship.
    
\noindent\textbf{Competing interests:} 
The authors declare that they have no competing interests.

\noindent\textbf{Data and materials availability:}  
CVSVGD Python API is available at: \url{https://github.com/ZhangGroup-MITChemistry/CVSVGD}

\noindent\textbf{Supporting Information:}
Technical Details of our RBF Fitting process, convergence plots for the alanine dipeptide system, repeat of figure \ref{fig:alanine_tripeptide} with 500 CV-space particles, convergence plots for umbrella sampling simulations, as well as videos of CV-space trajectories for the two toy model systems, the alanine dipeptide system and alanine tripeptide system.
\end{acknowledgement}

\clearpage
\newpage
\providecommand{\latin}[1]{#1}
\makeatletter
\providecommand{\doi}
  {\begingroup\let\do\@makeother\dospecials
  \catcode`\{=1 \catcode`\}=2 \doi@aux}
\providecommand{\doi@aux}[1]{\endgroup\texttt{#1}}
\makeatother
\providecommand*\mcitethebibliography{\thebibliography}
\csname @ifundefined\endcsname{endmcitethebibliography}  {\let\endmcitethebibliography\endthebibliography}{}

\definecolor{myblue}{RGB}{0, 123, 167}
\def\mcitetrackID{si}\def\mcitebibtrackID{si}
\setcounter{figure}{0}
\renewcommand{\thefigure}{S\arabic{figure}}
\setcounter{table}{0}
\renewcommand{\thetable}{S\arabic{table}}
\setcounter{equation}{0}
\renewcommand{\theequation}{S\arabic{equation}}
\renewcommand{\theHequation}{S\arabic{equation}}   

\renewcommand{\citenumfont}[1]{S#1}
\renewcommand{\bibnumfmt}[1]{(S#1)}

\clearpage
\newpage
\section{Supporting Information}

\subsection*{Fitting Radial Basis Functions to Mean-Forces}
We utilized the radial basis function (RBF) fitting method of Maragliano et al. \cite{maragliano_single-sweep_2008} to recover the free energy surfaces for the systems investigated. The fitting method is described in the following.
An RBF is of the following form:
\begin{equation}
    \varphi(\zeta)=\sum_{i=1}^{N}\alpha_ik\left(\zeta|\mu_i,\sigma\right),
\end{equation}
where $k(\cdot)$ is any suitable basis function, $\zeta=(\xi_1,\dots,\xi_d)$ is a vector of $d$ CV-space coordinates, $\alpha_i$ is a constant which assigns a weight to the $i$-th RBF component, $\mu_i$ is the center of the $i$-th RBF component, and $\sigma$ is a bandwidth parameter controlling the basis function's width. In our investigations we use the typical gaussian kernel for non-periodic CVs:
\begin{equation}
    k\left(\zeta|\mu,\sigma\right)=\exp\left[{-\frac{\left({\zeta-\mu}\right)^2}{2\sigma^2}}\right],
\end{equation}
and the von Mises kernel for periodic ones:
\begin{equation}
    k\left(\zeta|\mu,\sigma\right)=\exp\left[{\frac{1}{\sigma}\cos\left(\zeta-\mu\right)}\right]. 
\end{equation}
Given a set of mean forces $\{\nabla_\zeta F_i\}_{i=1}^N$, and corresponding force centers $\{\zeta_i\}_{i=1}^{N}$, we wish to parameterize our RBF $\varphi(\zeta)$ such that the following loss is minimized:
\begin{equation}
    \mathcal{L}=\sum_{i=1}^N\Vert\nabla_\zeta F_i-\nabla_\zeta\varphi(\zeta_i)\Vert^2
\end{equation}
The RBF's gradient $\nabla_\zeta\varphi(x)$ is a sum of the gradient of basis functions expanding as follows:
$$
\nabla_\zeta\varphi(\zeta_i)=\sum_{j=1}^N\alpha_j\nabla_\zeta k\left(\zeta_i|\zeta_j,\sigma\right)
$$
\begin{equation}
    \mathcal{L}=\sum_{i=1}^N\Vert\nabla_\zeta F_i-\sum_{j=1}^N\alpha_j\nabla_\zeta k\left(\zeta_i|\zeta_j,\sigma\right)\Vert^2
\end{equation}

The loss function can be then be equivalently rewritten as a matrix equation:
$$
\mathcal{L} = \left\Vert K\alpha-f\right\Vert^2
$$
Where $K$ is a $(dN)\times N$ matrix, where elements of matrix $K$ are given by 
$$
k_{ij}=\frac{\partial}{\partial\xi_{\lfloor (i-1)/N\rfloor+1}} k\left(\zeta_{((i-1)\bmod N)+1}|\zeta_j,\sigma\right)
$$
$\alpha=(\alpha_1,\dots,\alpha_N)$ is a vector of the weight parameters, and $$
f=\left(\frac{\partial}{\partial\xi_{\lfloor (1-1)/N\rfloor+1}}F_{((1-1)\bmod N)+1},\dots,\frac{\partial}{\partial\xi_{\lfloor (dN-1)/N\rfloor+1}}F_{((dN-1)\bmod N)+1}\right)
$$
is a vector containing all mean-forces component-wise, in other words a flattened $d\times N$ matrix of the combined mean-forces. 
To prevent over-fitting, we also employ regularization on $\alpha$ with a regularization parameter $\lambda>0$, giving the final loss of:
\begin{equation}
\mathcal{L} = \left\Vert K\alpha-f\right\Vert^2+\lambda\Vert\alpha\Vert^2
\label{eq:fitting_loss}
\end{equation}

A $\lambda=10^{-9}$ was used for all RBF fittings. The optimal solution to eq.\ref{eq:fitting_loss} can be found using reduced QR-factorization, where we factorize the augmented matrix $A$:
$$
A=\left[ \begin{array}{c} K \\ \sqrt{\lambda}I_N \end{array} \right]
$$
Where $I_N$ is an $N\times N$ identity matrix. Factoring $A=QR$, gives:
\begin{equation}
R\alpha=Q^Tb
\label{eq:qr_factor_eqn}
\end{equation}
Where:
$$
b=\left[\begin{array}{c}f \\ \overrightarrow{0} \end{array}\right]
$$
and $\overrightarrow{0}$ is a N-component 0-vector. Using eq. \ref{eq:qr_factor_eqn}, the optimal $\alpha$ can be found via back-substitution.
To ensure the optimal fit of all parameters $\alpha$ and $\sigma$, we use a Pytorch\cite{paszke_pytorch_2019} implementation of the described fitting procedure, allowing the $\sigma$ to be additionally optimized via gradient descent. This is done by initializing the fitting procedure with a guess $\sigma$ value, finding the optimal $\alpha$ given that $\sigma$ value, and then computing the gradient of eq. \ref{eq:fitting_loss} with respect to $\sigma$, and iterating until convergence. Notably, reduced QR-factorization is fully differentiable in the Pytorch implementation, enabling quick convergence to the optimal parameters. Additionally, it allows for anisotropic kernels via a unique $\sigma$ per CV dimension to better fit the free energy surface.

\clearpage
\newpage
\section*{Supplemental Figures}
\begin{figure*}{
    \centering
    \includegraphics[width=1.0\linewidth]{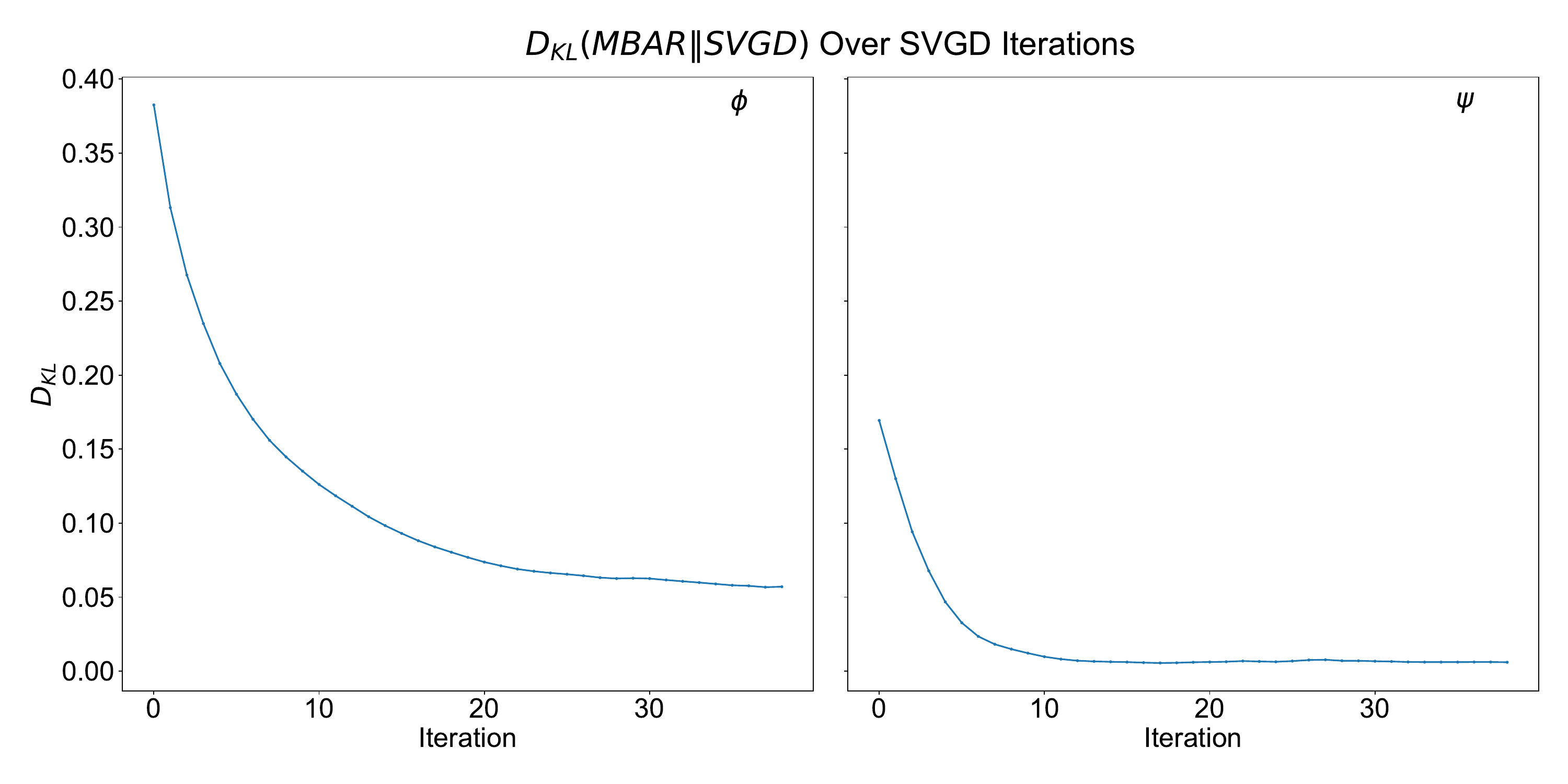}
\caption{
\textbf{Convergence of CV-SVGD particle distributions for aqueous-phase alanine dipeptide.}
Kullback--Leibler divergence ($D_{\mathrm{KL}}$) between kernel density estimates (KDEs) of the CV-SVGD particle distributions and the umbrella-sampling reference distributions as a function of SVGD iteration. The divergence plateaus after approximately 20 iterations for both collective variables, suggesting convergence of the particle distributions within this number of iterations.
}   
    \label{fig:si_alanine_dipeptide_conv}
} 
\end{figure*}
\begin{figure*}
    \centering
    \includegraphics[width=1.0\linewidth]{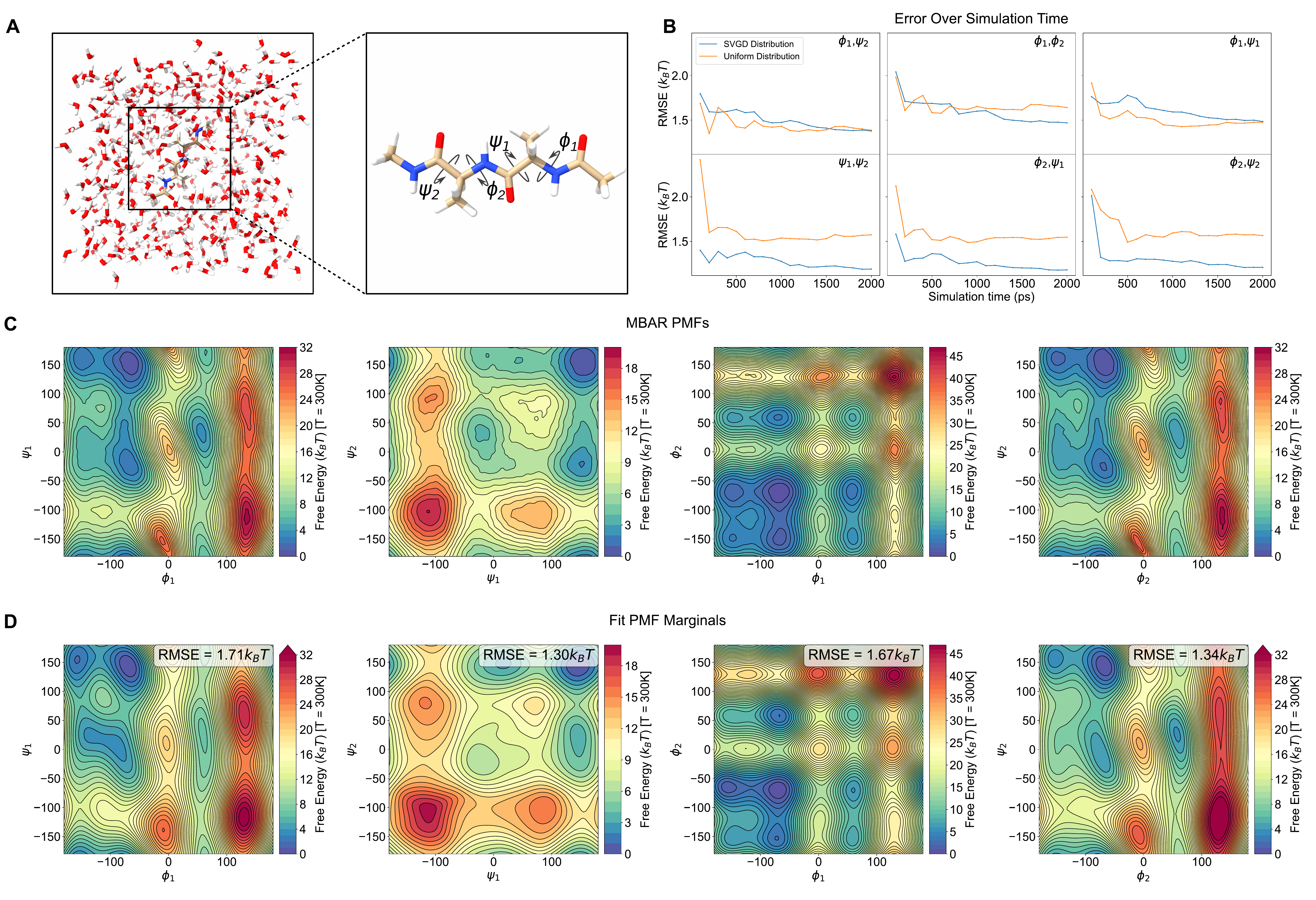}
\caption{
\textbf{Free-energy reconstruction of aqueous-phase alanine tripeptide using 500 CV-SVGD particles.}
Same analysis as in Figure~4, but with 500 CV-SVGD particles instead of 200. The free-energy surfaces shown in panel D were reconstructed using mean forces estimated from 250~ps of restrained MD simulation per particle.
}
    \label{fig:si_alanine_tripeptide}
\end{figure*}

\begin{figure*}{
    \centering
    \includegraphics[width=1.0\linewidth]{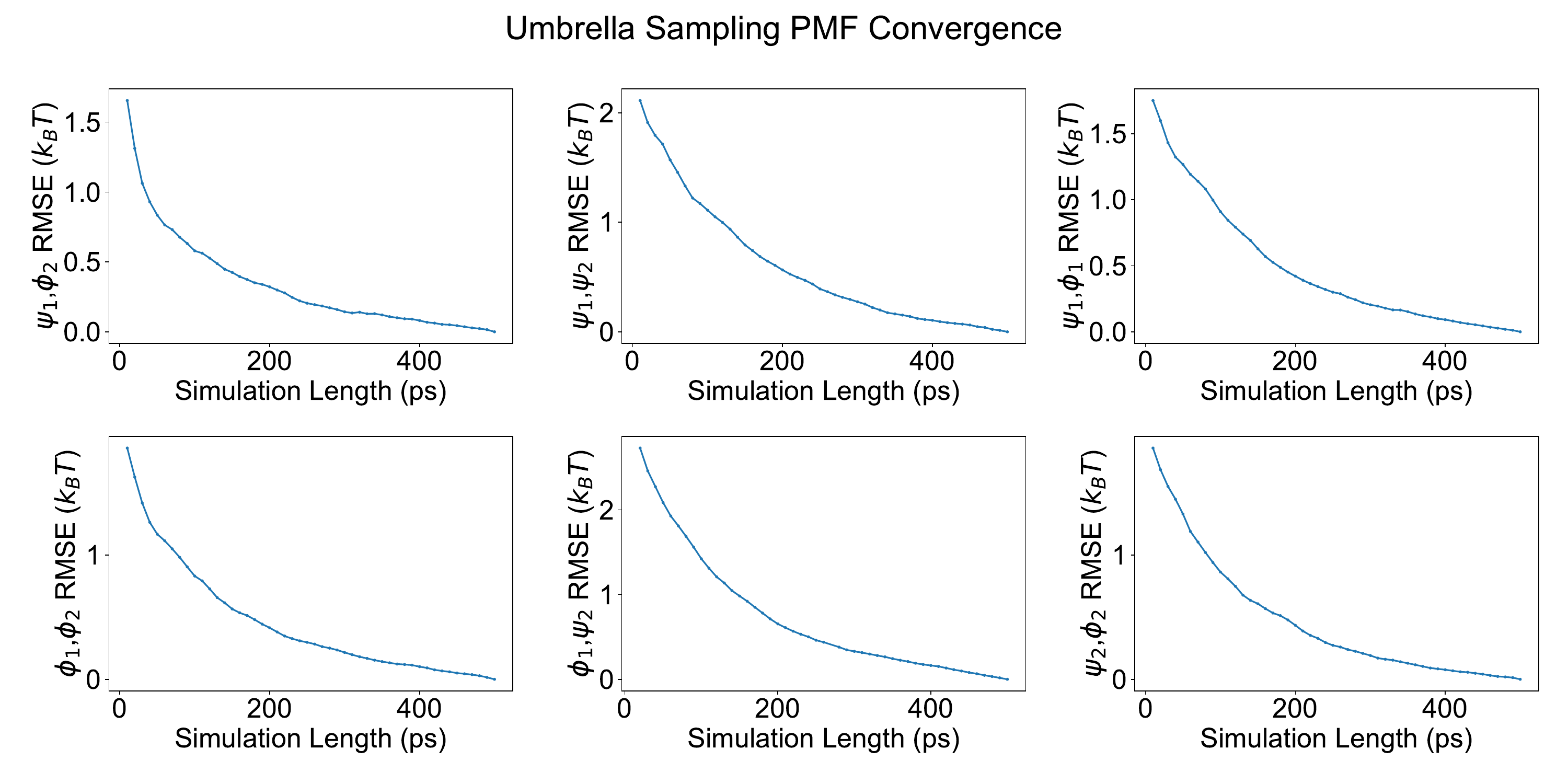}
\caption{
    \textbf{Convergence of umbrella-sampling free-energy profiles for aqueous-phase alanine tripeptide.}
    Root-mean-square error (RMSE) of the free-energy profiles as a function of simulation length for all six sets of two-dimensional umbrella-sampling calculations. The RMSE was calculated relative to the corresponding reference profile obtained using the full simulation trajectory. The pair of collective variables sampled in each calculation is indicated along the $y$ axis.
}
\label{fig:si_us_convergence}
}
\end{figure*}
\newpage
\clearpage
\makeatletter
\providecommand{\latin}[1]{#1}
\makeatletter
\providecommand{\doi}
  {\begingroup\let\do\@makeother\dospecials
  \catcode`\{=1 \catcode`\}=2 \doi@aux}
\providecommand{\doi@aux}[1]{\endgroup\texttt{#1}}
\makeatother
\providecommand*\mcitethebibliography{\thebibliography}
\csname @ifundefined\endcsname{endmcitethebibliography}  {\let\endmcitethebibliography\endthebibliography}{}

\end{document}